\documentclass[prb,reprint,amsmath]{revtex4-1}
\usepackage{graphicx}
\usepackage{dcolumn}
\usepackage{bm}
\usepackage{upgreek}
\newcommand{\unit}[1]{\,\text{#1}}

\begin{document}
\title[Why do bubbles in Guinness sink?]{Why do bubbles in Guinness sink?}
\author{E. S. Benilov}
 \altaffiliation[]{Department of Mathematics, University of Limerick, Ireland}
 \email{Eugene.Benilov@ul.ie}
 \homepage{http://www.staff.ul.ie/eugenebenilov/hpage/}
 \author{C. P. Cummins} \altaffiliation[]{Department of Mathematics,
   University of Limerick, Ireland} \email{Cathal.Cummins@ul.ie}
 \author{W. T. Lee} \altaffiliation[]{MACSI, Department of
   Mathematics, University of Limerick, Ireland}
 \email{William.Lee@ul.ie}
 \homepage{http://www.ul.ie/wlee/}

\begin{abstract}
Stout beers show the counter-intuitive phenomena of sinking bubbles
while the beer is settling. Previous research suggests that this
phenomena is due the small size of the bubbles in these beers and the
presence of a circulatory current, directed downwards near the side of
the wall and upwards in the interior of the glass. The mechanism by
which such a circulation is established and the conditions under which
it will occur has not been clarified. In this paper, we demonstrate
using simulations and experiment that the flow in a glass of stout
depends on the shape of the glass. If it narrows downwards (as the
traditional stout glass, the pint, does), the flow is directed
downwards near the wall and upwards in the interior and sinking
bubbles will be observed. If the container widens downwards, the flow
is opposite to that described above and only rising bubbles will be
seen.
\end{abstract}

\maketitle

\section{Introduction}

Stout beers, such as Guinness, foam due to a combination of dissolved
nitrogen and carbon dioxide,\cite{LeeDevereux11} as opposed to other
beers which foam due to dissolved carbon dioxide alone. The use of
nitrogen results in a range of desirable characteristics of the beer,
including a less bitter taste and a creamy long-lasting head which can
be attributed to the low solubility of nitrogen and small size of the
bubbles.\cite{Denny09,Bamforth04} This small bubble size is also
responsible, at least in part, for another intriguing characteristic
of stout beers: the phenomenon of sinking bubbles, observed while the
beer is settling, i.e.\ between the pouring of the beer and the
formation of the head.\cite{ZhangXu08}

Experimental studies\cite{AlexanderZare??} have demonstrated that
the phenomenon of sinking bubbles is real and not an optical
illusion, while simulations\cite{Service00} show that the
bubbles are driven by a downward flow, the velocity of which
exceeds the upward velocity of the bubble due to the Archimedean
force. The existence of such a flow near the wall of the glass
implies that there must be an upward flow somewhere in the
interior. The mechanism of this circulation is, however, unclear,
as is the role of the shape of the glass.

Understanding these types of bubbly flows is important for a number of
applications, such as manufacturing champagne glasses engraved with
nucleation sites,\cite{LigerEtAl07} widget and similar technologies
for promoting foaming in stouts,\cite{LeeEtAl11,Gerstner2011}
designing glasses which minimize the settling time of stouts and,
generally, for industrial processes involving bubbly flows
(e.g.\ bubble columns\cite{LapinEtAl02}).

In this paper, we put forward an explanation of the effect of sinking
bubbles in Guinness, which takes into account the role of the shape of
the glass. First, in Section~\ref{Properties} we describe the
properties of Guinness as a two-phase medium. In
Section~\ref{Simulations} we present the results of numerical
simulations for several shapes of the glass. In
Section~\ref{Mechanism} we explain the basic mechanism which drives
bubbles downwards and describe a simple experiment which can be used
to confirm our hypothesis. Finally we give conclusions in
Section~\ref{Conclusions}.

\section{\label{Properties} Properties of Guinness} 

We shall model Guinness by a liquid of density $\rho_{l}$ and
viscosity $\mu_{l}$, with randomly distributed bubbles of gas of
density $\rho_{g}$ and viscosity $\mu_{g}$. For a temperature of
$6^{\circ}\mathrm{C}$ (recommended for consumption of Guinness by its
producer \textquotedblleft Diageo\textquotedblright\cite{FAQ??}) and
normal atmospheric pressure, we have%
\begin{align*}
\rho_{l}&=1007\unit{kg}\unit{m}^{-3}
 & \mu_{l}&=2.06\times10^{-3}\unit{Pa}\unit{s}
\\
\rho_{g}&=1.223\unit{kg}\unit{m}^{-3}
 & \mu_{g}&=0.017\times10^{-3}\unit{Pa}\unit{s}
\end{align*}
where the former values have been measured by ourselves and
verified against the extrapolation formula of Ref.~\onlinecite{ZhangXu08}.

To check whether the bubble shapes differ from spheres, we introduce
the Bond number
\[
\mathrm{Bo}=\frac{\rho_{l}gd_{b}^{2}}{\sigma}
\]
where $d_{b}$ is the bubbles' characteristic diameter, $\sigma$ is the
surface tension of the liquid/gas interface, and $g$ is the
acceleration due to gravity. Assuming $d_{b}=122\unit{}\upmu\text{m}$
(as reported in Ref.~\onlinecite{RobinsonEtAl08}) and
$\sigma=0.745\unit{N}\unit{m}^{-1}$ (which corresponds to water/air
interface), we obtain $\mathrm{Bo} \approx0.002$ -- which is
sufficiently small to assume that bubbles in Guinness are spherical.

Note also that Guinness (as well as vast majority of `real'
liquids) contains a lot of surfactants, which make the bubbles
behave as rigid spheres.\cite{Batchelor00} This allows one to
estimate the characteristic bubble velocity $u_{b}$ using the
Stokes formula for a rigid sphere,
\[
u_{b}=\frac{\left( \rho_{l}-\rho_{b}\right) gd_{b}^{2}}{18\mu_{l}}%
\approx3.96\unit{mm}\unit{s}^{-1}.
\]
Estimating the corresponding Reynolds number%
\[
\mathrm{\operatorname{Re}}=\frac{\rho_{l}u_{b}d_{b}}{\mu_{l}}\approx0.24,
\]
confirms that the Stokes formula yields a
qualitatively correct value for $u_{b}$. Furthermore, given that
$u_{b}$ is much smaller than the speed of sound, the gas can be
treated as incompressible.

Finally, we introduce the void fraction $f$, i.e.\ the gas's share of
the volume of the liquid/gas mixture. For canned Guinness,
$f\approx0.05$ (see Ref.~\onlinecite{RobinsonEtAl08}), whereas for
draught Guinness served in pubs, $f\approx0.1$ (according to our own
measurements). Note, however, that, traditionally, bartenders first
fill, say, 80\% of the glass and wait until it has fully settled
(i.e.\ all the bubbles have gone out of the liquid into the foamy
head), after which they would fill the glass full. Thus, when Guinness
is served to the customer, the void fraction can be estimated as
$f\approx0.02$, which is the value used in this work.

\section{\label{Simulations}Numerical modeling of the liquid/bubble circulation}

To simulate flows in Guinness, we use the finite element model for
bubbly flows included in the COMSOL Multiphysics package. The model's
physical foundations are described in detail in
Ref.~\onlinecite{SokolichinEigenbergerLapin04}. In this model the
bubbles are assumed to be all of the same size. In view of the
problem's axial symmetry, the axi-symmetric version of the model is
used.

\begin{figure}
\includegraphics[width=83mm]{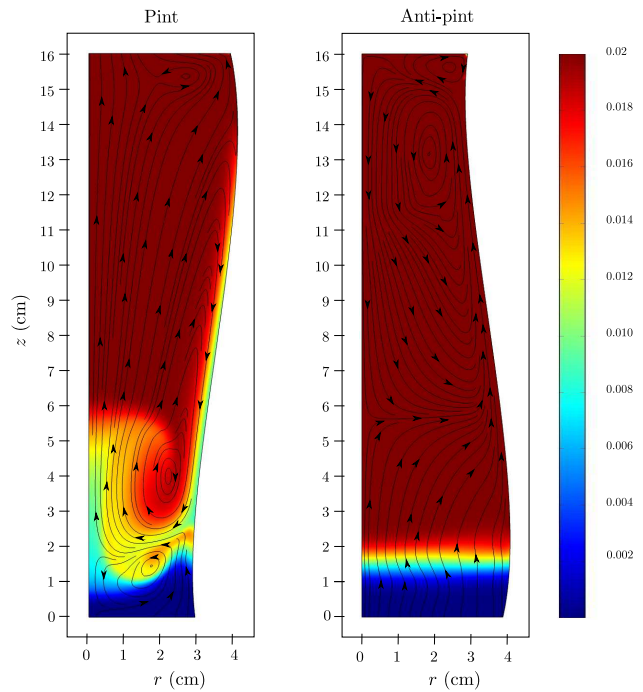}
\caption{Numerical simulations of bubbly flows for the pint and
anti-pint. The curves show the streamlines for the bubbles, the
color shows the void fraction $f$. The snapshots displayed
correspond to $t=4\unit{s}$. Observe the region of reduced
$f$ near the wall of the pint (the near-wall region of increased
$f$ in the anti-pint is not visible in this figure, but can be
observed in Fig.~\ref{fig2}).} \label{fig1}
\end{figure}

Two geometries of the holding container were examined (see
Fig.~\ref{fig1}): a pint and an `anti-pint', i.e. the pint turned
upside-down.  In both cases the initial distribution of bubbles was
uniform, and the physical parameters of Guinness were as described
above.

\begin{figure*}
\includegraphics[width=177mm]{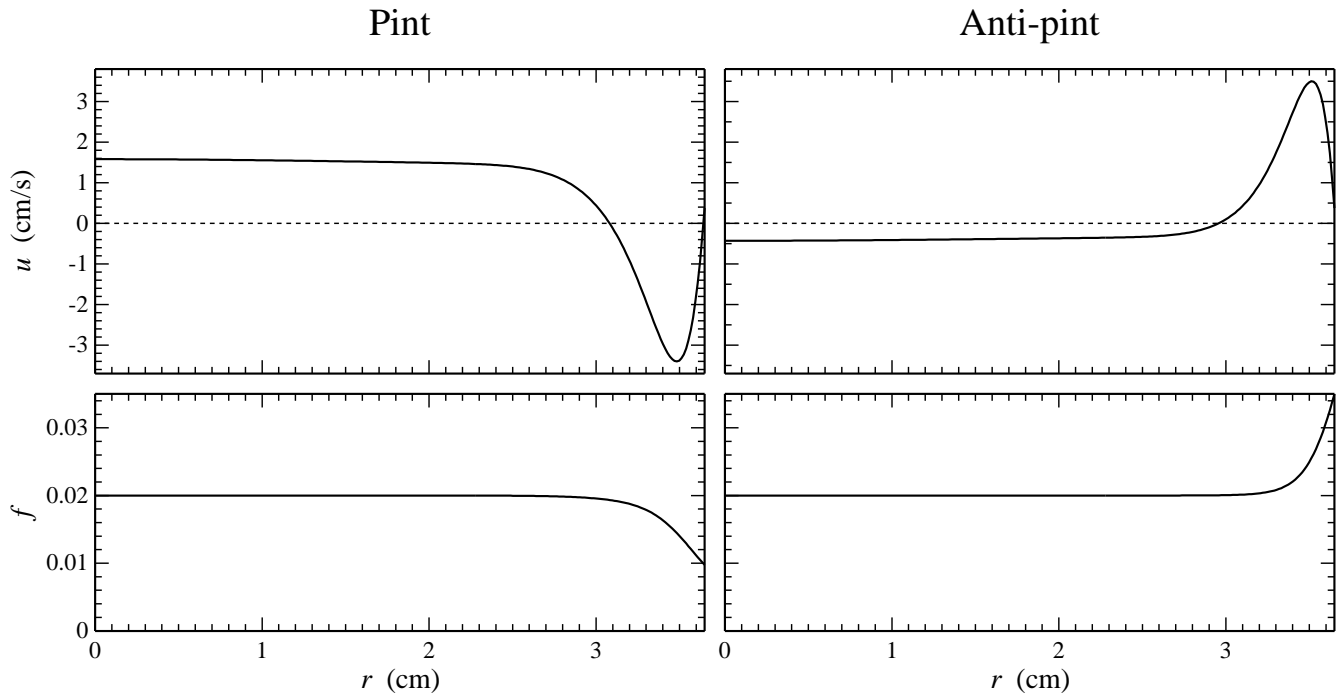}
\caption{The half-height cross-sections of the vertical velocity
$u$ and the void fraction $f$ for the pint and anti-pint
geometries (these graphs correspond to the $\left(  r,z\right)$
diagrams shown in Fig. 2). The dotted lines in the upper panels
separate the regions of upward/downward flow.} \label{fig2}
\end{figure*}

The results of typical simulations are shown in Fig.~\ref{fig1}. One can
see that an elongated vortex arises near the sloping part of the
pint container, resulting in a downflow of bubbles along the wall
(see the top-left panel of Fig.~\ref{fig2}). A similar vortex also exists
in the anti-pint, but it rotates in the opposite direction and,
thus, causes an \emph{upward} flow near the wall (see the
top-right panel of Fig.~\ref{fig2}).

Another important feature to be observed is the narrow region of low
density of bubbles along the wall of the pint container (see the left
panel of Fig.~\ref{fig1} and the lower-left panel of
Fig.~\ref{fig2}). In the anti-pint container, in turn, the bubble
density increases near the wall (which is not visible in the right
panel of Fig.~\ref{fig1}, but can be clearly seen in the lower-right
panel of Fig.~\ref{fig2}).

\begin{figure}
\includegraphics[width=83mm]{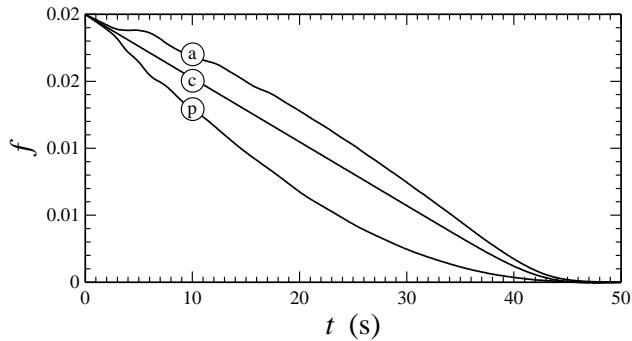}
\caption{The global void fraction $f$ (i.e.\ the proportion of the gas
  in the container) versus time $t$, for the three cases: the
  pint~(p), anti-pint~(a)---both illustrated in Fig.~\ref{fig2}---and
  a cylinder of the same volume~(c).} \label{fig3}
\end{figure}

We have also examined the evolution of the global void fraction for
the pint and anti-pint, as well as the cylindrical container of the
same volume. Fig.~\ref{fig3} shows that all three geometries provide
more or less the same settling time $T_{s}$ (for a glass of stout, a
smaller $T_{s}$ is generally regarded as an advantage).

$T_{s}$ has also been used to examine the extent to which our results
depend on the void fraction $f$ and the bubble size $d_{b}$ (these are
the only parameters with `uncertain' values).  It has turned out that,
surprisingly, the dependence of $T_{s}$ on $f$ is very weak: e.g., an
increase in $f$ from $0.02$ to $0.05$ results in a decrease in $T_{s}$
from $43$ to $46$ seconds (the settling time here was defined such
that $f(T_{s})=10^{-6}$). The dependence of $T_{s}$ on $d_{b}$ is much
stronger: a decrease in $d_{b}$ from $122$ to $90$ micron results in
an increase in $T_{s}$ from $43$ to $83$ seconds.

Note also that the settling time of, approximately, $43$ seconds
computed for $d_{b}=122\unit{}\upmu \text{m}$ and the pint container
does not agree with our experimental estimate of approximately $120$
seconds. The difference between the two results is probably caused by
the fact that all bubbles in our simulations were of the same size,
whereas in reality they are distributed with a certain dispersion. One
can then conjecture that smaller bubbles leave the liquid later than
those of the median size, which would account for the difference
between the computed and measured values of $T_{s}$.

In what follows, we shall argue that the circulation developing in
the flow is determined by the near-wall variation of the bubble
density (as suggested previously by Ref.~\onlinecite{Riviere99} for
bubble columns), and that the bubble density, in turn, is
determined by the shape of the container.

\section{\label{Mechanism} The mechanism of the effect} 

First of all, observe that, whichever way the bubbles move, they exert
a drag force on the surrounding liquid. This does not mean, however,
that the liquid is necessarily entrained by the motion of the bubbles.
Indeed, if, for example, the bubbles (and, hence, the drag force) are
distributed uniformly, all liquid particles must move the same way --
which effectively means that they cannot move at all due to the
liquid's incompressibility and the fact that the container has a
bottom. In this case, the drag force is compensated by a pressure
gradient exerted in the fluid.

Let us now assume that there is a region of low bubble density near
the container's wall (as there indeed is in the pint container). In
this case, the density of the drag force near the container's axis is
larger than that near the wall -- which creates an imbalance and,
thus, gives rise to a circulation: near the axis, the liquid flows
upwards and, near the wall, downwards.  Then, if the velocity of the
downward flow is larger than the relative velocity $u_{b}$ of the
bubbles, the bubbles will be observed to sink. A similar argument
indicates that a near-wall region with \emph{higher} bubble density
gives rise to an \emph{upward} flow (i.e. exactly what our simulations
show for the anti-pint container).

It still remains to identify the mechanism reducing the bubble
density near the wall for the pint geometry and increasing it for
the anti-pint one. Potentially, there are two such mechanisms. The
most obvious one is based on the `lift force', generated by the
flow around a sphere moving along a rigid boundary: this force
pushes the sphere away from the boundary
\cite{TakemuraMagnaudet09}. In the limit of small Reynolds number,
however, the lift force is weak, and estimates show that the
resulting reduction of the bubble density near the wall is
negligible. Another shortcoming of this mechanism is that it does
not seem to distinguish between the pint and anti-pint geometries.

\begin{figure}
\includegraphics[width=83mm]{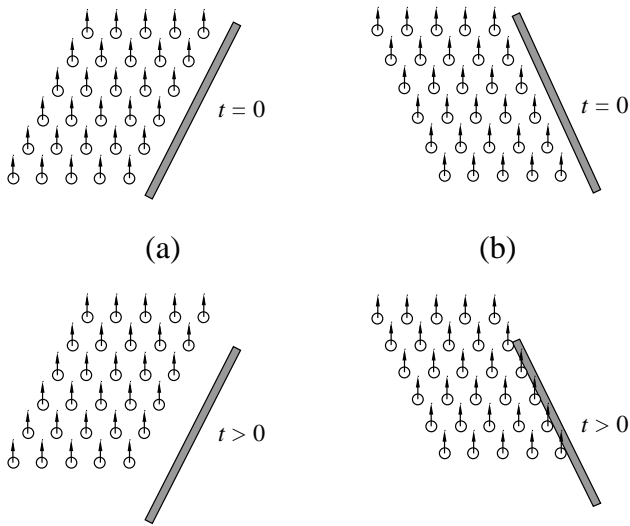}
\caption{The evolution of bubbles near the wall (a schematic), for
(a) container narrowing downwards (the bubbles move away from the
wall); (b) container widening downwards (the bubbles move towards
the wall and, eventually, accumulate there).} \label{fig4}
\end{figure}

To explore the effect of the geometry, assume that the container is
not cylindrical, but narrows slightly towards its bottom (as the pint
does). Then, even if the bubbles were initially distributed uniformly,
their upward motion immediately creates a bubble-free zone along the
wall (see Fig.~\ref{fig4}). On the other hand, in a container that
\emph{widens} towards its bottom (as the anti-pint does), the
initially upward motion of bubbles \emph{increases} the near-wall
bubble density. We believe that this simple kinematic effect is
responsible for the circulation observed in Guinness.

This effect, although not previously discussed in the context of
Guinness, is well known in sedimentation theory as the Boycott
effect.\cite{Boycott1920,Acrivos1979} It was first observed in test
tubes containing red blood cells when it was discovered that
sedimentation times could be significantly reduced by inclining the
test tubes.

Finally, our conclusions can be readily verified experimentally. If
Guinness is poured into a tall cylindrical container (such as, for
example, a laboratory measuring cylinder) and if the container is
tilted, bubbles will be observed to move upwards near its upper
surface and downwards near its lower surface -- in precise agreement
with the mechanism proposed (see supplementary material for a video of
this experiment).

\section{\label{Conclusions} Conclusions}

The sinking bubbles of Guinness and other stout beers have intrigued
beer drinking physicists and their students for some time. In this
paper we describe the role the shape of the Guinness pint glass plays
in promoting the circulatory flow responsible for the sinking
bubbles. In doing so we complete the explanation of this phenomena,
building on previous experimental and simulation work.  This allows us
to understand the physics underlying the shape of the Guinness pint
glass. It also raises the intriguing question: is this shape the most
efficient possible or could the settling time be significantly reduced
by some other, possibly non-axisymmetric, shape of pint glass?

\begin{acknowledgments}
The authors acknowledge the support of the Science Foundation
Ireland delivered through RFP Grant 11/RFP.1/MTH3281 and
Mathematics Initiative Grant 06/MI/005.
\end{acknowledgments}

\end{document}